\documentclass[aps,prl,twocolumn]{revtex4-2}
\usepackage{amsmath,bm}
\usepackage{graphicx,epsfig,braket,amssymb,amsfonts,eufrak,tabularx,multirow,amsmath,graphicx,color}
\usepackage{newtxtext}
\usepackage[colorlinks=true]{hyperref}
\usepackage[T1]{fontenc}
\usepackage{textcomp}

\newcommand{\bd}{\begin{displaymath}}
\newcommand{\ed}{\end{displaymath}}
\renewcommand{\vec}[1]{{\bf #1}}
\newcommand{\bpm}{\begin{pmatrix}}
\newcommand{\epm}{\end{pmatrix}}

\begin{document}

\title{Magnetoelastic Coupling–Driven Chiral Spin Textures: A Skyrmion–Antiskyrmion-Like Array}

\author{Gyungchoon Go}
\affiliation{Department of Physics, Korea Advanced Institute of Science and Technology, Daejeon 34141, Korea}

\author{Se Kwon Kim}
\affiliation{Department of Physics, Korea Advanced Institute of Science and Technology, Daejeon 34141, Korea}

\begin{abstract}
We theoretically demonstrate that sufficiently strong magnetoelastic coupling can change the ground state of otherwise uniform spin systems to chiral spin configurations. More specifically, we show that, a periodic array of chiral spin textures can spontaneously emerge in a two-dimensional ferromagnetic system on a substrate—even in the absence of Dzyaloshinskii–Moriya interaction. The resulting spin texture resembles a skyrmion–antiskyrmion lattice, characterized by alternating scalar spin chirality and a nonuniform but sign-preserving out-of-plane spin profile. Our analysis reveals that such patterns form naturally when the magnetoelastic interaction is sufficiently strong, while the coupling between flexural phonons and the substrate is sufficiently weak. These findings uncover a previously unexplored mechanism for chiral spin texture formation driven purely by magnetoelastic coupling, signaling at potential utilities of materials with strong magnetoelastic responses.
\end{abstract}

\maketitle


\emph{Introduction.}\textemdash
Chiral spin textures, such as chiral domain walls and skyrmions, have attracted significant interest in spintronics owing to their rich physics often associated with their topological nature and also because of their practical applications.
For example, a moving skyrmion is known to exhibit a Hall effect due to the topological Magnus force~\cite{Zang2011,Nagaosa2013, Everschor2014,Jiang2017}. When they form a periodic arrangement as a skyrmion crystal, they can give rise to a giant topological Hall effect, originating from the emergent electromagnetic field induced by their chiral spin structure~\cite{Bruno2004, Tatara2007, Neubauer2009, Kanazawa2011, Nagaosa2012}.
Because of their topological stability and nanoscale size, skyrmions are considered promising candidates for next-generation memory and logic devices~\cite{Fert2013, Tomasello2014, Zhang2015}.
Although skyrmions have been realized in a variety of materials, they have predominantly been confined to systems with noncentrosymmetric crystal structures and strong spin–orbit coupling~\cite{Rosler2006, Yi2009, Muhlbauer2009, Yu2010, Munzer2010, Han2010, Seki2012, Shibata2013}.
Such structural conditions enable the Dzyaloshinskii–Moriya interaction (DMI), which plays a central role in stabilizing chiral spin textures~\cite{Dzyaloshinskii1958, Moriya1960}.

Although DMI has been regarded as the principal mechanism for chiral magnetism,
such noncollinear spin structures can also be stabilized by other mechanisms such as long-range dipolar interaction~\cite{Lin1973,Takao1983,Lee2016,Montoya2017,Jiang2017a},
four spin-exchange interactions~\cite{Heinze2011}, and frustrated exchange interactions~\cite{Okubo2012,Leonov2015,Lin2016}.
However, the potential role of magnetoelastic coupling in stabilizing chiral spin structures has not been explored, which is at odd considering the universal presence of the magnetoelastic coupling in magnetic systems. This raises the question of whether spin–lattice interactions can drive the formation of chiral spin configurations in centrosymmetric systems where no DMI is present.

In this work, we predict that magnetoelastic coupling arising from single-ion anisotropy striction~\cite{Kittel1949, Kittel1958} can drive the otherwise uniform spin configurations into chiral spin textures.
We show that beyond a critical coupling strength, the uniform ferromagnetic ground state becomes unstable, leading to spin canting state accompanied by a distortion
of the flexural phonon amplitudes. As a result, in a one-dimensional system, a kink-soliton configuration can emerge. We further demonstrate that, in a two-dimensional magnetic layer supported by a substrate, a periodic array of chiral spin structures can spontaneously form when the magnetoelastic coupling is sufficiently strong.
The resulting spin texture takes the form of a skyrmion–antiskyrmion-like lattice, characterized by alternating scalar spin chirality and a nonuniform yet sign-preserving out-of-plane spin component.
This two-dimensional spin density wave (SDW) pattern emerges without invoking DMI, highlighting a purely spin–lattice-driven mechanism for chiral texture formation. Unlike previously studied skyrmion crystals, our skyrmion-like and antiskyrmion-like configurations carry non-quantized charges without topological protection. Our findings demonstrate that chiral spin textures can be realized in systems without DMI, based solely on magnetoelastic coupling, whereby establishing magnetoelastic coupling as a viable route to chiral spin textures in two-dimensional magnetic materials.

\emph{Model.}\textemdash
We begin with the spin Hamiltonian given by
\begin{align}
H_{\rm spin} = -J \sum_{\langle i,j \rangle} \vec S_i\cdot \vec S_j  - g \mu_B \sum_i \vec B_{\rm eff} \cdot {\vec S}_{i},
\end{align}
where $J>0$ denotes a ferromagnetic Heisenberg exchange and $\vec {B}_{\rm eff}$ is an effective magnetic field
that includes the external and anisotropy fields.
For simplicity we assume $\vec B_{\rm eff}$ points strictly along the $z$-axis, so that the uniform $S_z$-aligned state is the ground state.
We complement this with a harmonic lattice Hamiltonian of out-of-plane (flexural) phonon mode
\begin{align}\label{Hamph}
H_{\rm ph} = \sum_{i} \frac{({p}^z_i)^2}{2M} + \frac{K_Z}{2} \sum_{i,{\vec e}_i} ({u}_{i}^z - {u}_{i+{\vec e}_i}^z)^2,
\end{align}
where $u_i^z$ and $p_i^z$ are the $z$-component of the $i$th ion’s displacement and conjugate momentum, respectively,
and $K_Z$ is the spring constant.
To describe the interaction between lattice distortions and spin textures, we invoke the magnetoelastic coupling.
In this work, we assume the spins remain nearly aligned along the $z$-axis for simplicity, in which the coupling of longitudinal and transverse phonon modes ($u_x$ and $u_y$) to spins can be neglected over the coupling of flexural mode to spins in the magnetoelastic interaction. Then, the magnetoelastic Hamiltonian can be written as~\cite{Kittel1949, Kittel1958, Thingstad2019, Go2019}
\begin{align}
H_{\rm mp} &= \kappa  \sum_i {S_{i,z}} \sum_{{\vec e}_i} ({\vec S}_i \cdot {\vec e}_i)  (u^z_i - u^z_{i+ {\vec e}_i})
\end{align}
where $\kappa$ parametrizes the coupling strength and the inner sum runs over nearest-neighbor bond vectors~${\mathbf e}$.

\begin{figure}[t]
\includegraphics[width=1.0\columnwidth]{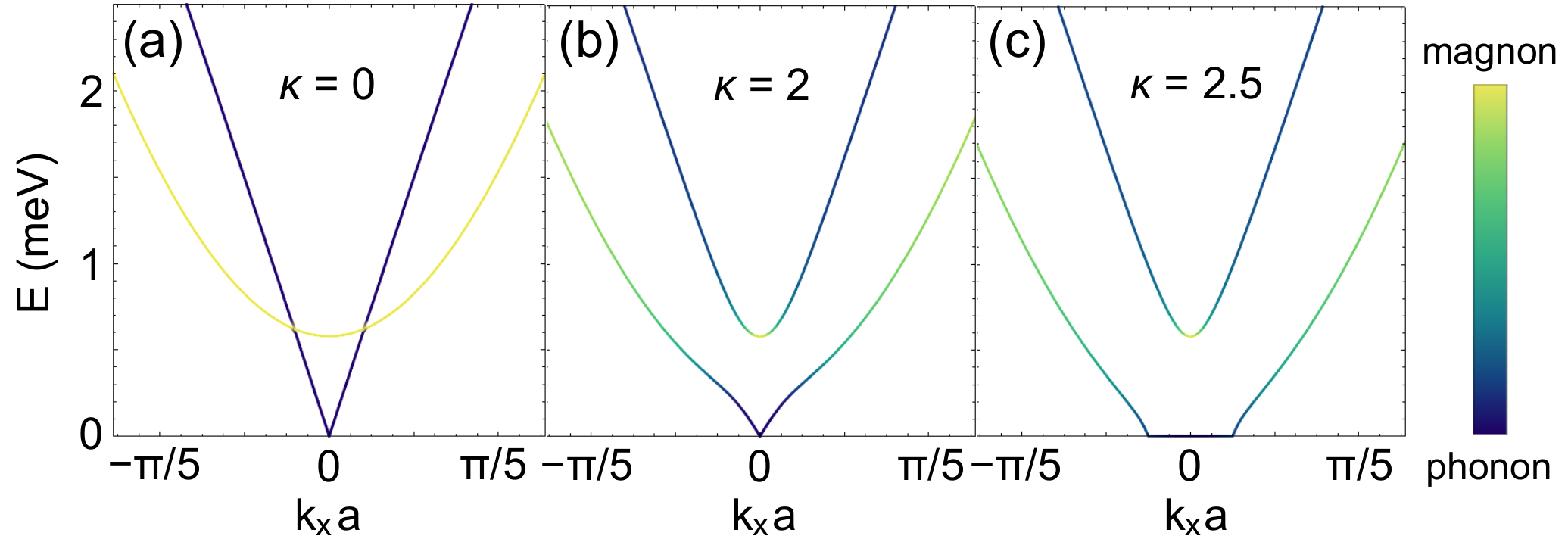}
\caption{Band structure for (a) $\kappa = 0$, (b) $\kappa = 2$, (c) $\kappa = 3$ meV/\AA.
Parameters used are: $S = 5/2$, $J = 1$ meV, $g = 2$, $B_{\rm eff} = 5$ T, $M = 55$ u, and $K_Z = 30$ eV/nm$^2$, yielding $\kappa_c \approx 2.34$ meV/\AA.}\label{fig:1}
\end{figure}

Figure~\ref{fig:1} shows the band structure of a two-dimensional square-lattice magnet computed via the Holstein–Primakoff approach,
assuming a uniform ground state with all spins aligned along $\vec {\hat z}$. The parameters used in the figure are as follows: $S = 5/2$, $J = 1$ meV, $g = 2$, $B_{\rm eff} = 5$ T, $M = 55$ u, and $K_Z = 30$ eV/nm$^2$.
For $\kappa < \kappa_c \approx 2.34$ meV/\AA, the magnon–phonon interaction opens a finite band gap without perturbing the ground state [Fig.~\ref{fig:1}(b)], where
$\kappa_c$ denotes the critical coupling strength.
However, for $\kappa > \kappa_c$, the uniform–ground-state assumption fails to produce the proper magnon dispersion (see the lower band in[Fig.~\ref{fig:1}(c)]), indicating that the magnon–phonon coupling should destabilize the assumed uniform ground state by distorting it.

\begin{figure}[t]
\includegraphics[width=1\columnwidth]{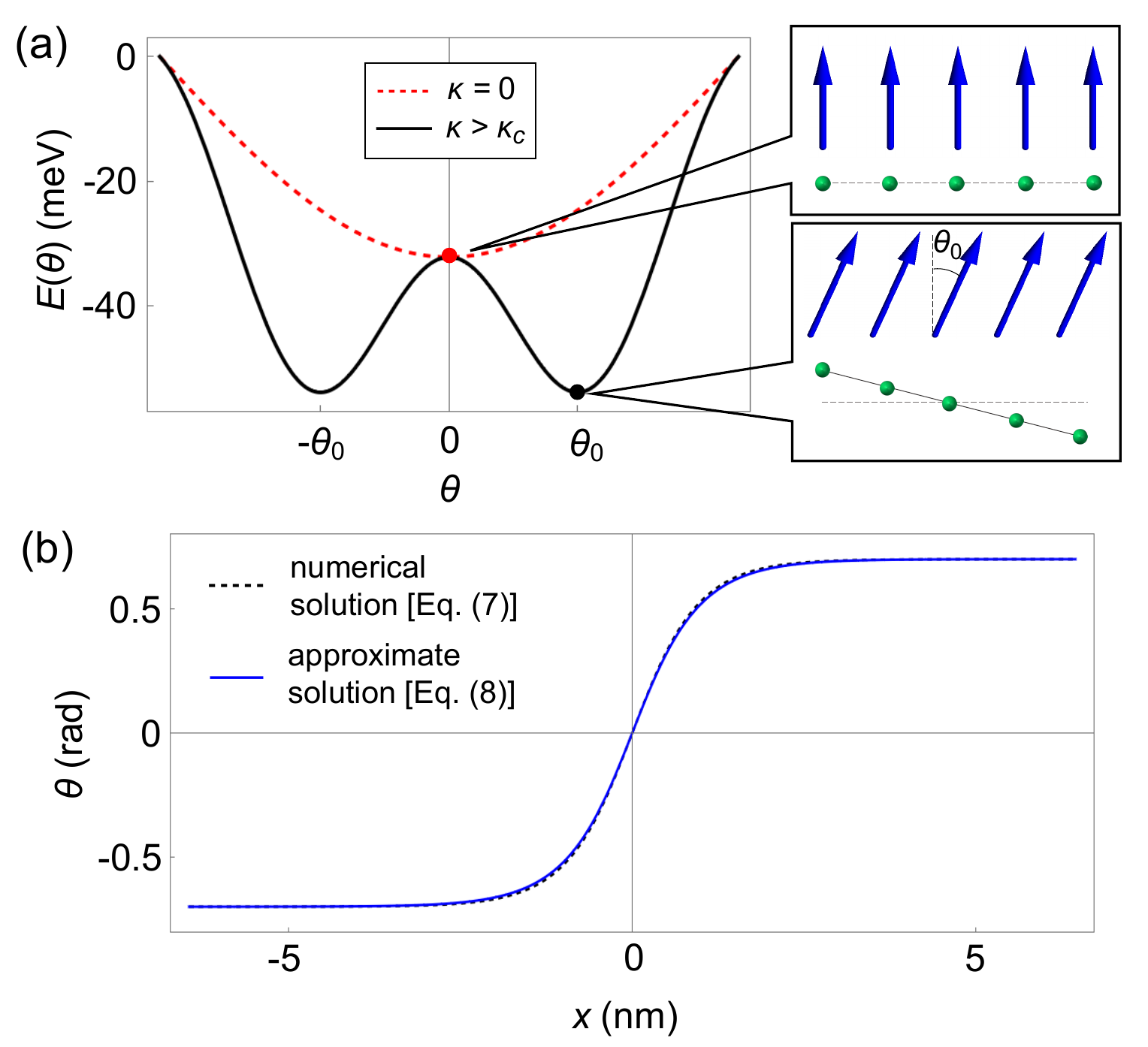}
\caption{(a) Effective potential profile of the 1D model for different magnetoelastic couplings.
For $\kappa > \kappa_c$, there are two degenerate energy minima, corresponding to tilted angle $\theta = \pm \theta_0$ (blue arrows),
accompanied by a uniform displacement gradient (green dots).
(b) Numerical (dashed) and approximate (solid) solutions interpolating between these two minima.
Parameters used are: $S = 5/2$, $J = 1$ meV, $g = 2$, $\kappa = 5$ meV/\AA, $B_{\rm eff} = 5$ T, $K_Z = 30$ eV/nm$^2$, and $a = 3$ \AA.}\label{fig:2}
\end{figure}

\emph{Spin reorientation due to the magnon-phonon interaction.}\textemdash
To intuitively illustrate how magnon–phonon coupling modifies the ground state,
we first consider a one-dimensional chain model along the $x$-axis,
subject to free boundary conditions, representing an ideal, substrate-free system.
In the continuum limit, the energy of the system can be written as
\begin{align}\label{conE}
E(x) = A ({\vec m}')^2 - b s_z + C (u_z')^2 + \tilde\kappa m_x m_z u_z',
\end{align}
where $\vec m (x)$ is the unit vector field representing the local magnetic order,
${\vec m}' (=\partial_x \vec m)$ and $u_z' (= \partial_x u_z)$ denote spatial derivatives with respect to $x$, and $A = {2 J S^2}/{a}$, $b = {g \mu_B B_{\rm eff} S}/{a^3}$, $C = {K_Z}/{a}$, $\tilde\kappa = {2 \kappa S^2}/{a^2}$ with a lattice constant $a$.
Substituting ${\vec m} = (\sin\theta \cos\phi, \sin\theta\sin\phi, \cos\theta)$
into the energy functional, then carrying out the functional variation with respect to $\phi$ and $u_z'$ yields
$\phi = 0$ $(m_y =0)$ and $u_z' = -\tilde\kappa \sin(2\theta)/(4C)$ and
\begin{align}
E(\theta) =  A\theta'^2 +  V(\theta),
\end{align}
where $V(\theta) = -\frac{\tilde\kappa^2}{16 C} \sin^2 (2\theta) - b \cos\theta$.
From the effective potential profile $V(\theta)$ shown in Fig.~\ref{fig:2}(a),
we find that once the magnetoelastic coupling exceeds the critical value $\tilde\kappa_c = \sqrt{2 b C}$,
the ground state configuration develops a uniform tilt $\theta = \pm\theta_0$, where
\begin{align}
\theta_0 = \cos^{-1}\left(\frac{2 + 3 {\cal F}^{2/3}}{6{\cal F}^{1/3}}\right),
\end{align}
with ${\cal F} = \frac{2\tilde \kappa_c^2}{\tilde \kappa^2} + \sqrt{\left(\frac{2\tilde \kappa_c^2}{\tilde \kappa^2}\right)^2 - \left(\frac{2}{3}\right)^3}$
and a concomitant,
spatially uniform strain (a constant displacement gradient) $u_z' \propto \sin(2\theta_0)$.
This implies that, although this modulation increases the energies of both the spin and lattice subsystems,
the resulting decrease in magnetoelastic energy dominates at sufficiently strong coupling, leading to a net reduction in the total energy~[see Fig.~\ref{fig:2}(a)].
Because the system has two energetically degenerate vacua at $\theta = \pm\theta_0$,
one can construct a soliton by interpolating between these two minima, subject to the Euler–Lagrange equation
\begin{align}\label{ELeq}
\theta'' +  \frac{\tilde\kappa^2}{16 A C}\sin4\theta - \frac{b}{2A} \sin\theta = 0.
\end{align}
As the effective potential is too complex to admit a closed-form solution, we employ the small-angle approximation.
In this limit, Eq.~\eqref{ELeq} reduces to sine-Gordon equation: $\theta'' + g \sin(4\theta) = 0$ with $g = [\tilde\kappa^2/(2C) - b]/(8A)$,
whose kink solution is $\theta(x) = \tan^{-1}\left[\tanh(\sqrt{g}(x-x_0))\right]$, This solution connects the distinct vacua $\pm \pi/4$, which do not coincide with the true vacua $\pm \theta_0$.
Enforcing the exact boundary values, $\theta(\pm \infty) = \pm \theta_0$, by a simple scaling factor gives the approximate soliton profile
\begin{align}\label{Apsol}
\theta_\pm(x) \approx \pm\frac{4\theta_0}{\pi} \tan^{-1}\left[\tanh({\sqrt g}(x-x_0)\right].
\end{align}
Figure~\ref{fig:2}(b) compares the approximate soliton profile [Eq.~\eqref{Apsol}] with the numerical solution of the Euler–Lagrange equation [Eq.~\eqref{ELeq}].

\begin{figure}[t]
\includegraphics[width=1.0\columnwidth]{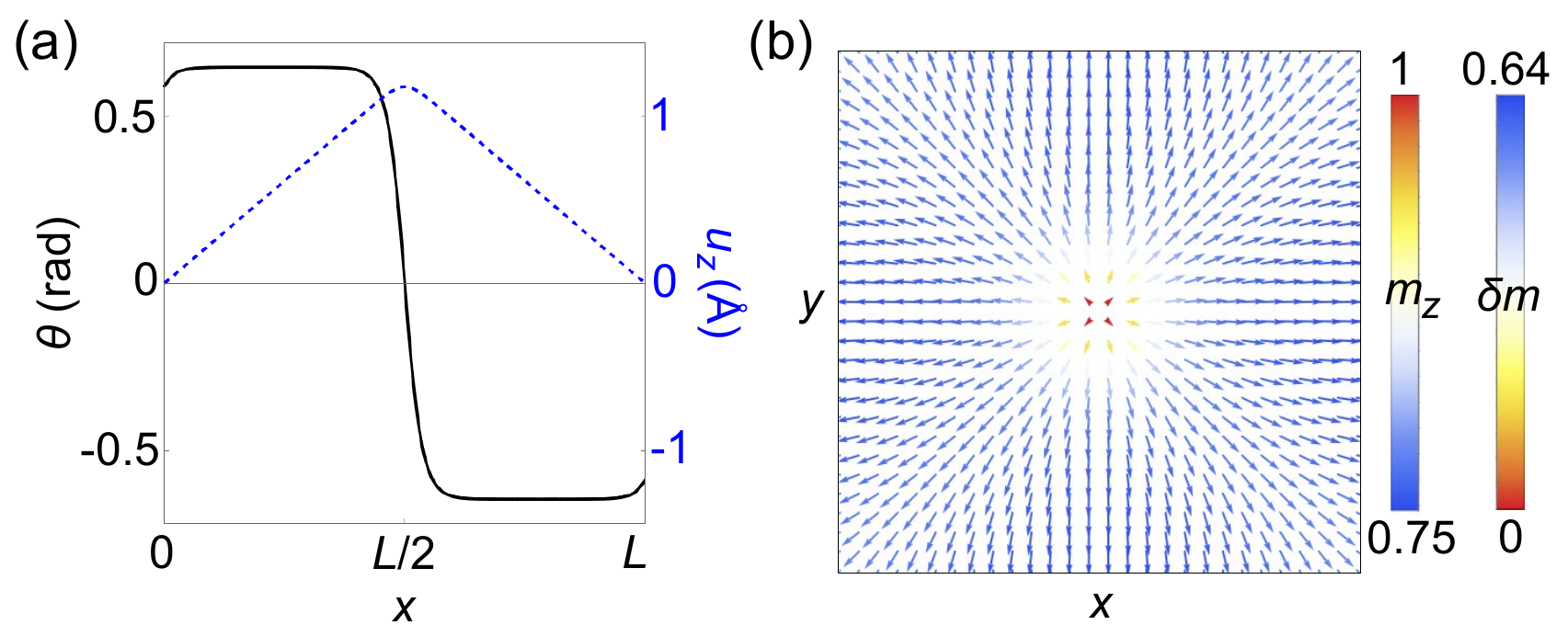}
\caption{(a) Simulated lowest-energy configuration of the 1D model with length $L = 15$~nm.
(b) Simulated lowest-energy configuration of the 2D model obtained on a square lattice
(only the central $26\times26$ region of the full $50\times50$ lattice is shown for clarity).
$\delta m = \sqrt{1 - m_z^2}$ represents magnitude of in-plane spin component.
In both cases, a fixed boundary condition is imposed on the lattice displacement, while a free boundary condition is applied to the spin vector.
Parameters used are: $S = 5/2$, $J = 1$ meV, $g = 2$,  $M = 55$ u, $\kappa = 4$ meV/\AA, $B_{\rm eff} = 2$ T, $K_Z = 30$ eV/nm$^2$, and $a = 3$ \AA.
Details of the simulation method are discussed in {\it Chiral spin textures}.}\label{fig:3}
\end{figure}

While the idealized 1D model under free boundary condition provides intuitive insight into the tilted ground state $[\theta(x) = \pm\theta_0]$,
the resulting constant displacement gradient $[u_z' \propto \sin(2\theta)]$ leads to a divergence of $u_z$ in the thermodynamic limit. To avoid this unphysical situation, we impose a fixed boundary condition on $u_z$ (i.e., $u_z = 0$ at boundaries) and a free boundary condition on $\theta$, mimicking a suspended membrane that is clamped at its boundaries.
Under this condition, the lowest-energy configuration in 1D chain model involves one kink ($\theta_+$) or anti-kink ($\theta_-$) soliton,
and the lattice displacement exhibits a non-sinusoidal, half-cycle mode with wavelength $\lambda = 2L$, increasing and then decreasing (or vice versa) across the system~[see Fig.~\ref{fig:3}(a)].
In contrast, an excited state forms a one-dimensional SDW characterized by an alternating array of kink and anti-kink solitons.

Extending the one-dimensional chain model [Eq.~\eqref{conE}] to two dimensions, we obtain the same energy minima $\theta =\theta_0$, with an arbitrary constant phase $\phi$.
In this case, a domain-wall or anti-domain-wall configuration, which leads to kink soliton extending along one-dimensional lines, no longer represents a minimum-energy state.
Instead, a point-like soliton--formed by the intersection of kinks in both the $x$ $(\phi = 0)$ and $y$ $(\phi = \pi/2)$ directions--is preferred,
as this configuration maximizes the area of minimum energy state ($\theta \approx \theta_0$) compared to the one-dimensional domain wall.
In this case, the azimuthal spin angle $\phi$ winds by $2\pi$ around the point-like soliton, producing a skyrmion (or antiskyrmion) like spin texture
with non-integer topological charge~[see Fig.~\ref{fig:3}(b)].

The divergence of $u_z$ due to the lattice distortion can also be remedied by introducing a uniform on-site potential for $u_z$, representing the effect of a substrate support.
Although this additional term precludes a analytic solution, its qualitative effect is clear:
it suppresses a uniform, runaway distortion of $u_z$.
Instead, the system stabilizes a SDW featuring periodic modulations of both $u_z$ and $m_{x,y}$.
Thus, in two-dimensions, the spin texture evolves into a chiral pattern characterized by an alternating skyrmion–antiskyrmion-like array.
It is worth noting that the resulting chiral spin texture exhibits a skyrmion density distribution.

\begin{figure}[t]
\includegraphics[width=1.0\columnwidth]{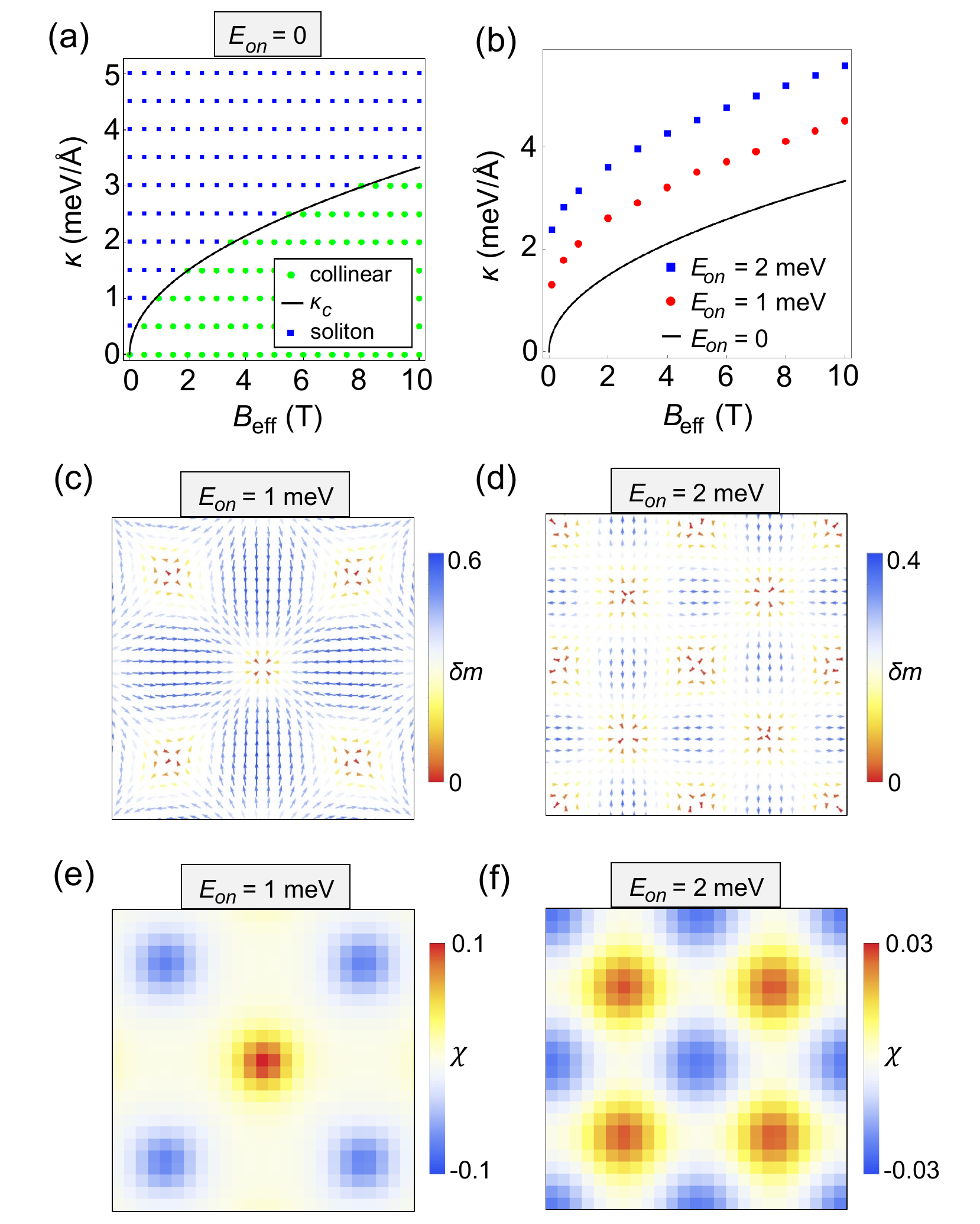}
\caption{(a) Ground-state phase diagram in the absence of on-site potential.
Green dots indicate the collinear spin phase and blue squares denote the single skyrmion-like soliton state.
(b) Critical magnon–phonon coupling $\kappa_c$ as a function of effective magnetic field for different on-site potential strengths.
The solid (black) line shows the analytical result at $E_{\rm on} = 0$, while dots and squares represent numerical results at $E_{\rm on} = 1$ meV and $E_{\rm on} = 2$ meV, respectively.
(c, e) Spin and scalar spin chirality textures of the 2D SDW state for $E_{\rm on} = 1$ meV, $\kappa = 0.04$, and $B_{\rm eff} = 2$ T.
(d, f) Same as (c, e), but for $E_{\rm on} = 2$ meV.
In (c–f), only the central $26\times26$ portion of the full $50\times50$ lattice is displayed for clarity.
}\label{fig:4}
\end{figure}

\emph{Chiral spin textures.}\textemdash
To verify the emergence of the chiral spin texture induced by magnetoelastic coupling,
we compute the energy-minimizing configuration of the total Hamiltonian $H = H_{\rm spin} + H_{\rm ph} + H_{\rm mp}$
using numerical simulations based on the arrested Newton flow method~\cite{Speight2020,Leask2022,Lee2024}.
In this approach, we solve a Newton's equation of motion of a ``particle" subject to the potential energy $E_{\rm tot} = E_{\rm spin} + E_{\rm ph} + E_{\rm mp}$.
Explicitly, we solve
\begin{align}\label{EOM}
\ddot{\Phi} = - \frac{\delta E_{\rm tot}[\Phi]}{\delta \Phi}, \quad \Phi = \{\vec S, u_z\},
\end{align}
with initial configuration of field $\Phi(0) = \Phi_0$ and initial velocity $\dot\Phi(0) = 0$.
The flow naturally evolves toward lower energy.
Equation~\eqref{EOM} is integrated using the fourth-order Runge–Kutta method, and the total energy is evaluated at the end of each time step.
If an increase in energy is detected, i.e., $E_{\rm tot} (t + \delta t) > E_{\rm tot} (t)$, the motion is halted by resetting the velocity: $\dot\Phi(t + \delta t) = 0$.
To ensure confidence in locating the global energy minimum, the algorithm is initialized with multiple distinct initial configurations.
The simulations are performed on a $50\times 50$ square lattice
using the following parameters: $S = 5/2$, $J = 1$ meV, $g = 2$, $B_{\rm eff} \in [0,10]$ T, $M = 55$ u, and $K_Z = 30$ eV/nm$^2$, and $\kappa \in [0,5]$ meV/\AA.
In the numerical analysis, fixed boundary conditions are imposed on the lattice displacement and free boundary conditions are applied to the spin angles.

Figure~\ref{fig:4}(a) shows the ground-state phase diagram in the absence of on-site potential $(E_{\rm on} = 0)$.
We observe that the ground state exhibits a single skyrmion-like soliton state shown in Fig.~\ref{fig:3}(b)
when the magnon-phonon coupling exceeds the critical value $\kappa_c$.
In Fig.~\ref{fig:4}(b), we present the magnetic field dependence of the critical coupling $\kappa_c$ for different on-site potentials,
which reflect the effect of substrate support on the flexural phonon mode $u_z$.
In Figs.~\ref{fig:4}(c) and (d), we show the 2D SDW configurations, which correspond to the lowest-energy states.
We note that a finite on-site potential suppresses lattice modulation and shortens the SDW period.
To visualize the spin chiral texture, we define the scalar spin chirality at site $i$ as $\chi_i = \chi^A_i + \chi^B_i$,
where $\chi^A_i = \vec S_i\cdot (\vec S_{i+ {\hat {\vec x}}}\times \vec S_{i + {\hat {\vec y}}})$ and $\chi^B_i = \vec S_i\cdot (\vec S_{i- {\hat {\vec x}}}\times \vec S_{i- {\hat {\vec y}} })$.
The resulting scalar spin chirality textures are shown in Figs.~\ref{fig:4}(e) and (f).
We note that the finite on-site potential induces spatially alternating scalar spin chirality
and a nonuniform yet strictly positive out-of-plane spin component, which are represented by a two-$\vec q$ modulation with mutually orthogonal wave vectors.
As expected, the wavelength of the modulation decreases as the on-site energy increases.
This is our main result: with sufficiently strong magnon–phonon coupling, a magnetic layer on a substrate stabilizes a two-dimensional chiral spin-density wave
featuring a skyrmion–antiskyrmion-like pattern.

\begin{figure}[t]
\includegraphics[width=1.0\columnwidth]{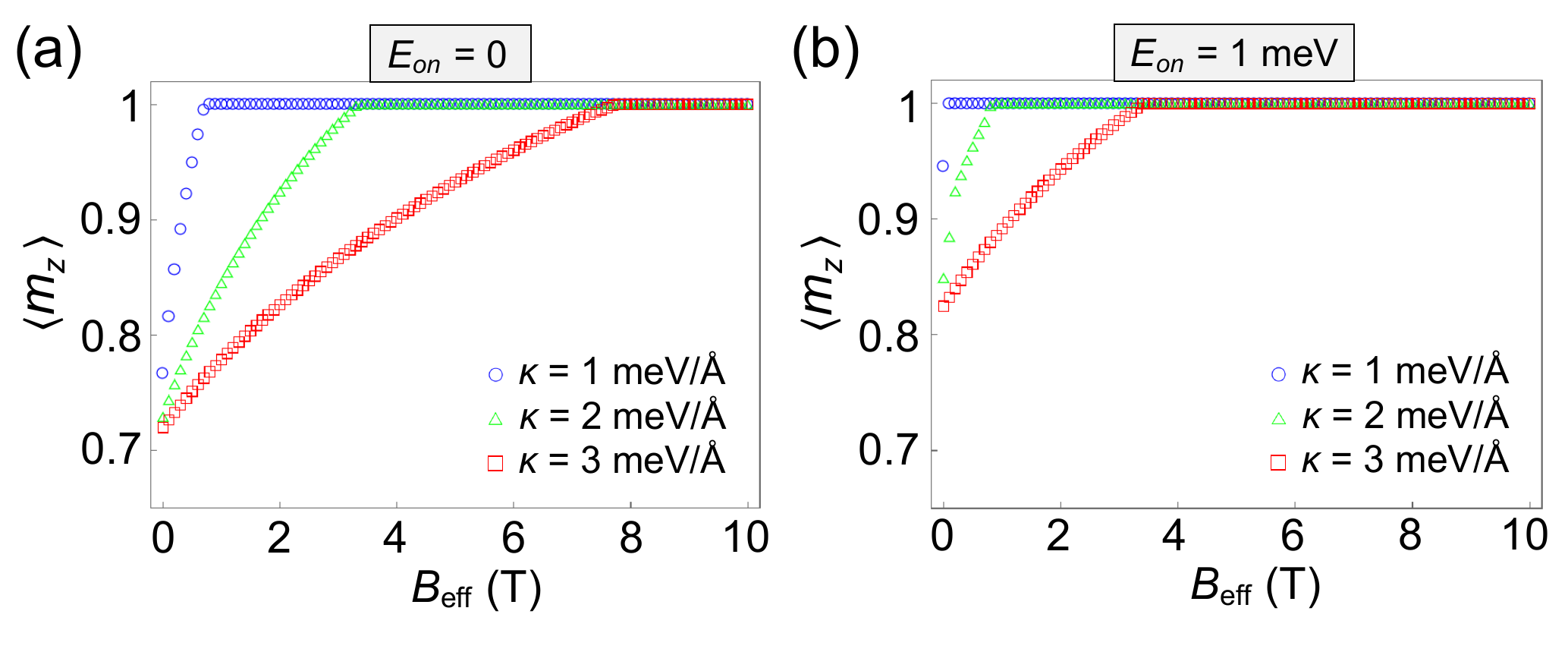}
\caption{Average value of $m_z$ as a function of $B_{\rm eff}$ in the lowest-energy state:
(a) without on-site potential ($E_{\rm on} = 0$),
(b) with finite on-site potential ($E_{\rm on} = 1$ meV).
}\label{fig:5}
\end{figure}

To assess the stability of the chiral SDW under external fields,
we show in Fig.~\ref{fig:5} the average out-of-plane magnetization $m_z$ as a function of magnetic field for the lowest-energy configuration.
As the field increases, the chiral SDW gradually evolves into a collinear magnetic state.
The field range over which the SDW is stabilized becomes broader with increasing $\kappa$, and narrower as the on-site potential $E_{\rm on}$ increases.

\emph{Discussion.}\textemdash
In this work, we theoretically demonstrate that magnetoelastic coupling can give rise to nontrivial spin textures in magnetic systems that have otherwise uniform spin configurations.
We show that, once the coupling strength exceeds a critical threshold, the uniform ferromagnetic ground state becomes unstable,
leading to spin canting and distortion of the flexural phonon modes.
Consequently, a kink-soliton configuration can form in a one-dimensional system.
In a two-dimensional ferromagnet, a periodic array of chiral spin structures can spontaneously emerge when substrate support effects are taken into account.
Unlike conventional skyrmion lattices that rely on the Dzyaloshinskii–Moriya interaction and exhibit topologically protected charges, the chiral textures identified here arise purely from magnetoelastic interactions and lack topological protection.

We note that the formation of such a SDW pattern requires sufficiently strong magnetoelastic coupling and weak interaction with the substrate.
Such a strong magnetoelastic coupling has been reported in metallic magnetic layers, including Co thin films,
with the magnetoelastic coefficient $\kappa$ estimated to be on the order of meV/\AA~\cite{GUTJAHRLOSER2001}.
However, since Co thin films are not intrinsically two-dimensional magnets, they are typically deposited on a substrate, where substrate-induced suppression of lattice modulations inhibits the emergence of SDW patterns.
In 2D magnets, magnetoelastic properties remain scarcely explored, and, to the best of our knowledge,
reliable values of $\kappa$ have rarely been reported to date,
with the exception of a density functional theory study on 2D antiferromagnets that estimates $\kappa \sim$ meV/\AA~\cite{Bazazzadeh2021}.
Several studies have indicated strong magnetoelastic coupling in certain 2D magnetic materials.
For example, density functional theory suggests that monolayer 2D magnets such as
Fe$_3$GeTe$_2$~\cite{Zhuang2016}, CrI$_3$~\cite{Staros2022}, and CrSBr~\cite{Xu2022} exhibit substantial magnetoelastic coupling.
Experimetal studies also report that Cr$_2$Ge$_2$Te$_6$ possesses strong magnetoelastic coupling~\cite{Carteaux1995,Sun2018,Tian2016,Siskins2022}.
The weak coupling to the substrate can be achieved using suspended membrane techniques for 2D magnets~\cite{Jiang2020,Siskins2022}.
Even when supported on hexagonal boron nitride (h-BN), the flexural (ZA) mode of graphene exhibits no discernible energy gap at the $\Gamma$-point~\cite{Pak2016}.
This suggests that h-BN interacts only weakly with the out-of-plane phonon modes,
implying that 2D magnets placed on h-BN substrates may also retain sufficiently soft flexural modes—thereby
satisfying the weak substrate interaction condition required for our mechanism.

Our theoretical study reveals magnetic soliton structures stabilized by magnetoelastic coupling,
featuring a spin configuration distinct from those previously reported in theory or experiment.
These results demonstrate that spin–lattice interactions alone can generate chiral spin textures,
highlighting unexplored opportunities in materials exhibiting strong magnetoelastic effects.

\section*{Acknowledgement}

We thank Seungho Lee for useful discussions.
G.G. acknowledges support by the National Research Foundation of Korea (NRF-2022R1C1C2006578).
S.K.K. was supported by the Brain Pool Plus Program through the National Research Foundation of Korea funded by the Ministry of Science and ICT (2020H1D3A2A03099291) and Samsung Science and Technology Foundation (SSTF-BA2202-04).

\bibliography{reference_ME}

\end{document}